\documentclass[floatfix,twocolumn,amsmath,amssymb,aps,prl,showpacs,letter]{revtex4}
\usepackage{latexsym,epsfig,bm,times,psfrag,subfigure}
\newcommand{\ag}{\boldsymbol{\alpha}}
\newcommand{\bg}{\boldsymbol{\beta}}
\newcommand{\cg}{\boldsymbol{\gamma}}
\newcommand{\dg}{\boldsymbol{\delta}}   

\begin{document}
\title{Random Walks and Anderson Localisation in a Three-Dimensional Class C Network Model}
\author{M. Ortu\~no, and A. M. Somoza}
\affiliation{Departamento de F\'isica, Universidad de Murcia,
  Murcia 30.071, Spain}
\author{J. T. Chalker} 
\affiliation{Theoretical Physics, Oxford University, 1, Keble Road, Oxford, OX1 3NP, United Kingdom}
\date{\today} 
\begin{abstract}
We study the disorder-induced localisation transition in a
three-dimensional network model that belongs to symmetry class C. The model
represents quasiparticle dynamics in a gapless spin-singlet
superconductor without time-reversal invariance. It is a special feature
of network models with this symmetry that the
conductance and density of states can be expressed as averages in a
classical system of dense, interacting random walks. 
Using this mapping, we present a more precise numerical study of
critical behaviour at an Anderson transition than has been possible
previously in any context.
\end{abstract}


\maketitle

Anderson transitions between diffusive and localised phases of quantum
particles in disordered systems constitute an important category of critical
phenomena. Depending on dimensionality and the symmetries of the
Hamiltonian, various universality classes are possible for
scaling behaviour \cite{review}. In most cases there is no known connection between
these Anderson universality classes and those for phase
transitions in classical systems, but in a special instance, known as
class C, properties of suitably chosen quantum lattice models can be
expressed in terms of observables for a classical model defined on the
same lattice. This mapping was originally discovered in
the context of the spin quantum Hall effect, where it relates a
delocalisation transition in two dimensions to classical
percolation, also in two dimensions, for which many relevant aspects of critical behaviour are known
exactly \cite{gruzberg}. In this paper we apply the mapping to a three-dimensional
system for which the classical counterpart is a model of interacting
random walks \cite{beamond}. The classical model is of interest both as
a representation of the quantum problem and in its own right.
While no exact results for its behaviour are available, the mapping 
makes possible simulations of the Anderson transition with unprecedented precision.

The symmetry class for localisation that we are concerned with is one of seven
recognised about a decade ago as being additional to the
three Wigner-Dyson classes that were originally identified in
the context of random matrix theory \cite{altland-zirnbauer}. Systems in these
additional symmetry classes are distinguished from ones in the
Wigner-Dyson classes by having a special point in the energy
spectrum and energy levels that appear in pairs either side of this
point. In particular, models in class C arise from the Bogoliubov
de-Gennes Hamiltonian for quasiparticles in a gapless, disordered spin-singlet
superconductor with broken time-reversal symmetry for orbital motion
but negligible Zeeman splitting. Here the special energy is the chemical potential
(which we set to zero) and pairs of levels are related by
particle-hole symmetry. Quasiparticle states in a three-dimensional
system of this type have an Anderson transition as a function of
disorder strength, which must be probed via spin or thermal transport
since charge transport is short-circuited by the
condensate. Particle-hole symmetry has profound consequences for
the influence of disorder on quasiparticle eigenstates and for 
the Anderson transition, which have been explored previously using
random matrix theory \cite{altland-zirnbauer}, the non-linear sigma
model \cite{senthil,zirnbauer}, and via calculations for one-dimensional systems \cite{one-d}. Most importantly,
the density of states at the chemical potential plays the role of an
order parameter for the transition, being finite in the metal and zero in the
insulator; by contrast, the density of states shows no critical
behaviour at transitions in the Wigner-Dyson classes.

The quantum to classical mapping provides a framework within which 
quasiparticle properties can be studied in great detail starting from a 
simplified description of a disordered superconductor.
It is based on a formulation of the
localisation problem as a network model \cite{network} in which
quasiparticles propagate along the directed links of a lattice and
scatter between links at nodes. Disorder enters the model in the form
of quenched random phase shifts associated with propagation on
links. For versions of the network model belonging to class C, first introduced in
Ref.~\cite{kagalovsky}, the disorder-averaged quasiparticle density of
states and (spin) conductance can be expressed as averages over
configurations of  interacting classical random walks on the same
directed lattice \cite{gruzberg,beamond}. This relation between
quantum properties and classical walks holds on any graph in which
all nodes have exactly two incoming and two outgoing links.
A single parameter $p$ controls behaviour at nodes:
incoming and outgoing links are arranged in pairs, and a particle passing through the node
follows the pairing with probability $p$, or switches with probability
$1-p$.
More precisely, in the quantum
problem the probabilities calculated by squaring amplitudes in the scattering matrix for the
node take the values $p$ and $1-p$, while in the classical problem the connection at each node between ingoing
and outgoing links is a quenched random variable having two possible arrangements with
these probabilities. Given a directed graph with the required coordination, 
any choice of classical connections at the
nodes separates paths on the graph into a set of distinct, closed,
mutually avoiding
walks. 
Average properties of these walks are calculated from a sum over
all node configurations, weighted according to their probabilities.
 
To use the mapping for a three-dimensional model one must select a four-fold
coordinated lattice, assign directions to the links so that
two are incoming and two are outgoing at every node, and pair incoming
with outgoing links. We pick the diamond lattice and choose link
directions and pairings at nodes 
so that the system has localised
states at $p=0$ and extended states at $p=1$. The resulting unit cell
contains 24 sites and we defer a full descrition until later in this
paper. We study samples consisting of $L\times L \times L$ unit cells
with two alternative sets of boundary conditions: using the classical mapping,
we calculate conductance between two opposite, open faces (with
periodic boundary conditions in the other directions), and we calculate the density of states in
closed samples with periodic boundary conditions in all three
directions.

The results of the mapping are as follows \cite{beamond}. First, the disorder-averaged spin
conductance $G(p,L)$ of the quantum system is given (in units of $\hbar/4\pi$) by
the average of the number $N_{\rm L}$ of classical paths from a
specified open face to the other. Second, the average density 
of states $\rho(\varepsilon)$ for eigenphases of a unitary time
evolution operator, which occupy the range $-\pi \leq \varepsilon \leq
\pi$ and play the role of energy levels, is given in terms of the
probability $P(s,p)$ in the classical problem for a given link to
belong to a closed walk of length $s$ steps, by
\begin{equation}
\rho(\varepsilon,p) = (2\pi)^{-1}\left[1- \sum_{s>0} P(s,p) \cos(2s\varepsilon)\right]\,.
\label{dos} 
\end{equation}
In the following we present and analyse data for $G(p,L)$
and $P(s,p)$, and deduce from the latter the critical properties of
$\rho(\varepsilon,p)$. In addition, we study the mean square end-to-end distance $\langle
R(s)^2\rangle$ of walks as a function of number of steps $s$. 
This last quantity has no direct significance for the original quantum
problem but provides a useful characterisation of the classical
walks.

We can study much larger systems using the classical
description than is possible for a quantum Hamiltonian with disorder: 
we use sizes in the range $40 \leq L \leq 440$, and calculate
averages over $10^7$ realisations for the smallest samples and
$4\times 10^3$ for the largest. The biggest systems hence contain over
$2 \cdot 10^9$ sites in total, and over $10^6$ sites in
cross-section. By contrast, direct calculations of eigenfunctions
using sparse matrix techniques \cite{evers,romer,markos} are limited to
systems $10^2$ or $10^3$ times smaller, while high precision studies
of Anderson transitions \cite{slevin} using
transfer matrix methods \cite{transfer} are restricted to systems
with a cross-section of around $10^3$ sites. Our system sizes enable us
to determine the values of critical exponents with about an order of
magnitude higher precision that has been possible numerically for
other Anderson transitions.

\begin{figure}
\includegraphics[width=.48\textwidth]{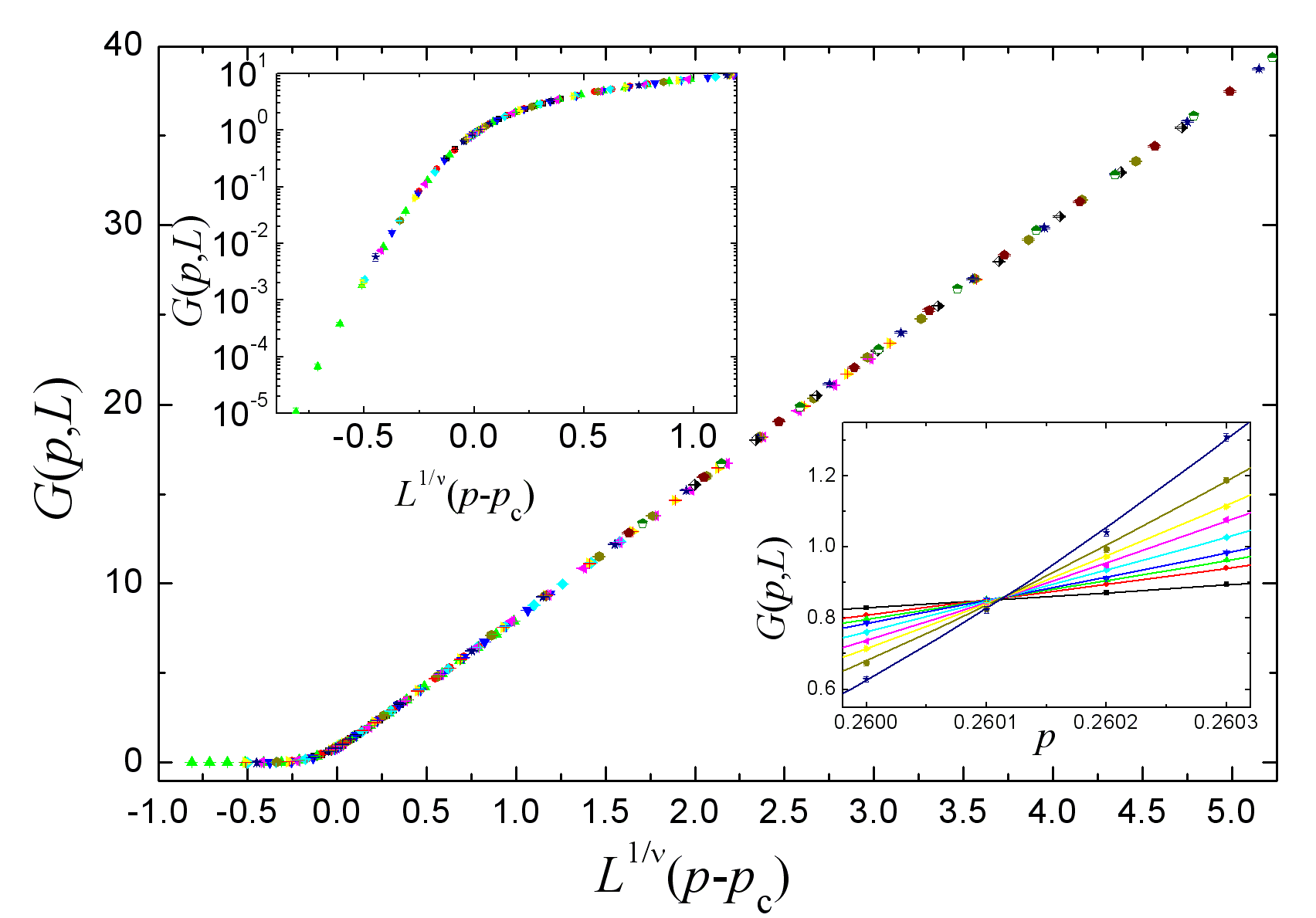}
\caption{(Color online) Conductance as a function of $(p-p_c)L^{1/\nu}$, illustrating
  scaling collapse. Upper inset: same data on a logarithmic conductance
  scale. Lower inset: conductance as a function of $p$ for several
  values of $L$. Lines are scaling fit described in main text.
}
\label{fig2}
\end{figure}
We discuss first our data for conductance $G(p,L)$. 
Our measurements span the range $0.25\leq p \leq 0.28$ and yield
values for $G(p,L)$ in the metal with statistical uncertainties of less than
$0.1\%$, except for the largest sample size ($L=440$) where they are $0.4\%$.
The model is designed to have a critical point at $p=p_{\rm c}$ separating
a metallic phase for $p>p_{\rm c}$ from an insulating one for
$p<p_{\rm c}$. From Ohm's law the metallic phase is characterised by
the behaviour $G(p,L) \sim \sigma(p)L$ for sufficiently large $L$, where $\sigma(p)$
is the conductivity, while in the insulator $G(p,L)\to 0$ at large $L$.
Close to the critical point we expect one parameter scaling in the form
\begin{equation}
G(p,L) = f(L/\xi(p))\,,
\label{scaling}
\end{equation}
where $\xi(p)$ is the localisation length in the insulator and the
correlation length in the metal. Hence curves of $G(p,L)$
as a function of $p$ for different $L$ intersect at $p_{\rm c}$, as
illustrated in the lower inset to Fig.~\ref{fig2}. In addition,
assuming that $\xi(p)$ diverges at the transition as $\xi(p) \propto
|p-p_c|^{-\nu}$, data for $G(p,L)$ in the critical regime
should collapse onto a single curve when plotted as a function of
$x\equiv (p-p_{\rm c})L^{1/\nu}$. This collapse is shown in Fig.~\ref{fig2}
using a linear scale for $G(p,L)$, and in the upper left inset using a
logarithmic scale to display more clearly behaviour in the insulator.


Our estimates for the values of $\nu$ and $p_{\rm c}$ obtained
from such a scaling analysis are
\begin{equation}
\nu=0.9985 \pm 0.0015
\end{equation}
and
$p_{\rm c} = 0.260116 \pm 0.000002 $.
The analysis in detail is as follows. Extending Eq.~(\ref{scaling})
to include corrections to scaling, we expect
\begin{equation}
G(p,L)= \tilde{f}(L^{1/\nu} \phi, L^{y_{\rm irr}} \psi)\;,
\end{equation}
where $\phi$ is the relevant scaling variable and $\psi$ is the
scaling variable for the leading correction, which is irrelevant
provided $y_{\rm irr} <0$. Since $G(L,p)$ varies smoothly with $p$ at fixed $L$,
the scaling variables have Taylor expansions 
\begin{equation}
\phi = (p-p_{\rm c}) + A(p-p_{\rm c})^2 \ldots
\label{corrections}
\end{equation} 
and $\psi = 1 + B(p-p_{\rm c}) \ldots$, in which leading coefficients
have been fixed via the definition of $\tilde{f}(x,y)$. As a first step, we omit all
corrections to scaling, setting $A$ and higher coefficients to
zero and considering $\tilde{f}(x,0)$. We construct this scaling
function using cubic B-splines. We judge all fits by comparing the
value of $\chi^2$ with the number of degrees of freedom. For the splines
we use 18 internal knot points at a spacing of $0.1$ for $-0.5\leq x
\leq 1.0$, where curvature of $\tilde{f}(x,0)$ is highest, and a spacing of
$1$ for $1 \leq x \leq 5$, where $\tilde{f}(x,0)$ is almost linear in $x$.
A coarser mesh of knot points does not allow an
adequate fit to the data, while a finer one does not significantly
improve the fit. Next we allow for an irrelevant scaling variable, by 
Taylor expanding $\tilde{f}(x,y)$ in $y$: we
find that the form
\begin{equation}
\tilde{f}(x,y) = \tilde{f}(x,0)[1+Cy]
\end{equation}
with $y_{\rm irr} = -1$ and $C$ constant provides an adequate
description of the data. Finally, we allow for a non-linear scaling
variable, as in Eq.~(\ref{corrections}), finding from the effect on
$\chi^2$ that non-zero $A$ is justified but non-zero $B$
or further terms are not.

Taking data with $p\leq p_{\rm max} = 0.273$ (excluded values are far from the critical
point) and $x\geq -0.5)$ (excluded values are deep in the insulator,
where errors in $G(L,p)$ are large), we have 262 points to fit and 26
parameters (the values of $\nu$, $p_{\rm c}$, $A$, $C$, 18 internal
spline points and four boundary ones). Our fit has $\chi^2 = 236.4$ and
is insensitive to the value of $p_{\rm max}$ in the range $0.265\leq
p_{\rm max} \leq 0.28$. Since the fit yields a value for $\chi^2$
close to the number of degrees of freedom, $262-26=236$, we believe
that errors in our values for $\nu$ and $p_{\rm c}$ are due mainly to
statistical errors in the data. We place 95\% confidence limits on these values
in two independent ways. In one we find the variation that increases $\chi^2$ by
4. In the other we use a Monte Carlo technique to generate synthetic
data \cite{numerical-recipes}. The two methods give the same results.

Our value for $\nu$ is in striking, although presumably accidental,
agreement with the leading order result from an $\epsilon$ expansion 
in $2 +\epsilon$ dimensions \cite{senthil}, $\nu=1/\epsilon$,
evaluated at $\epsilon=1$. It
is also consistent with the value  $\nu \approx 0.9$ found in an
earlier transfer matrix calculation \cite{previous} for a three-dimensional class C
network model, which did not use the mapping to classical random walks
and so was restricted to much smaller system sizes. The closeness of
our value to $\nu=1$ is tantalising but we are not aware of any theoretical
reason to attach significance to this. Indeed, the upper critical
dimension for the localisation transition in class C is believed to be
four \cite{andreas}, and so non-trivial exponent values are expected
in three dimensions.

\begin{figure}
\includegraphics[width=.48\textwidth]{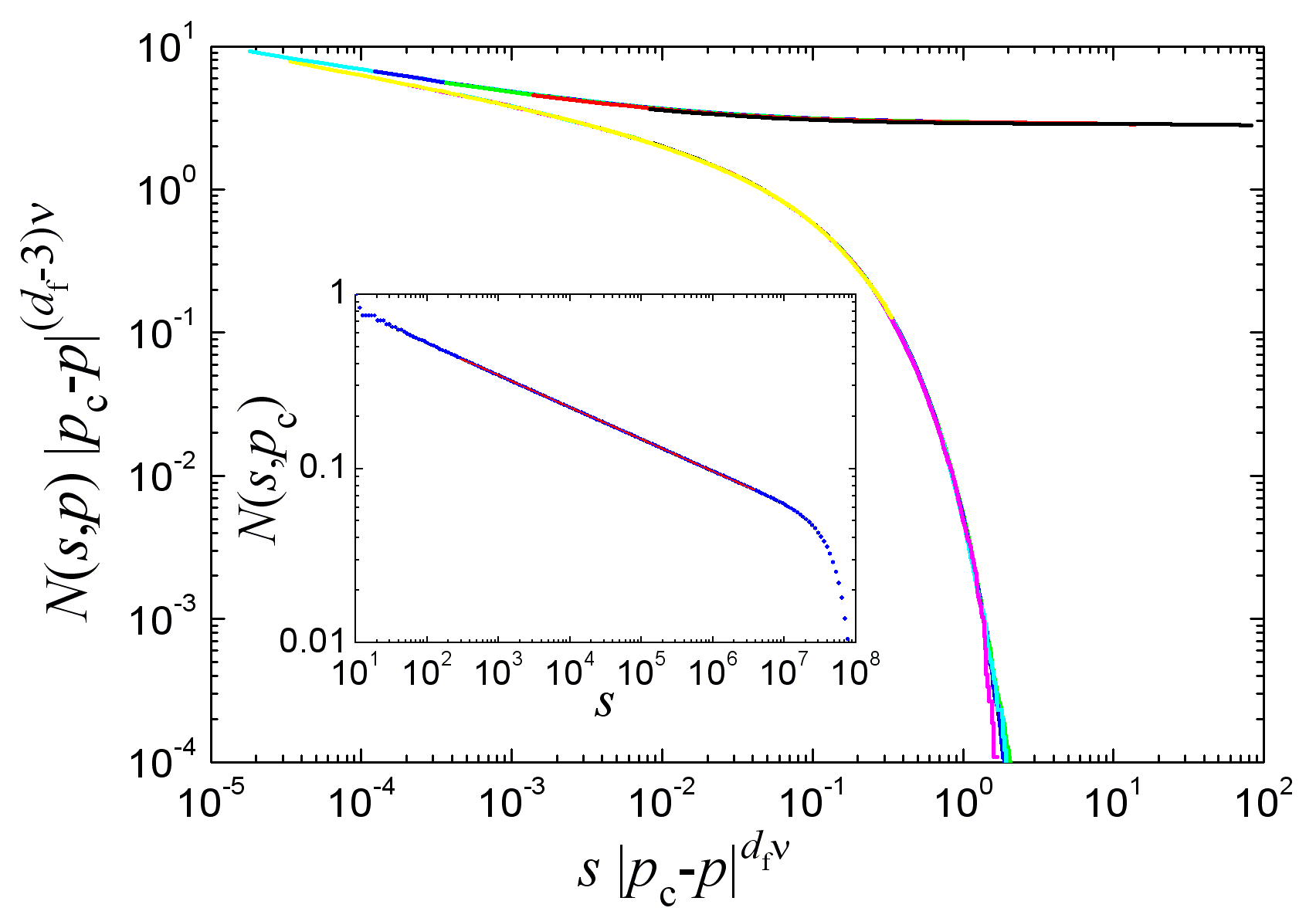}
\caption{(Color online) $|p-p_{\rm c}|^{(d_{\rm f} -3) \nu}N(s,p)$ as a function of $s|p-p_{\rm
    c}|^{d_{\rm f}\nu}$ on double logarithmic scales for several values of $p$ either side of $p_{\rm c}$. 
Inset: $N(s,p_{\rm c})$ as a function
of $s$ on double logarithmic scales. }
\label{fig4}
\end{figure}

We next present our data for the integrated return probability
$N(s,p)$, which is related to $P(s,p)$ via
\begin{equation}
N(s,p) = \sum_{t\geq s} P(t,p)\;.
\end{equation}
From this we
deduce the critical behaviour of the classical walks
and of the quantum density of states.
These data are obtained in closed samples by generating all walks for a given realisation of
node configurations and averaging over different
realisations, using sample sizes 
$L\leq 300$.
In the critical regime walks are fractal with dimension $d_{\rm f}$.
The length $\xi(p)$ sets their characteristic size in the insulator,
so that the characteristic arc length is $[\xi(p)]^{d_{\rm f}}$; in the
metal $\xi(p)$ represents a crossover length beyond which scaling properties
are as for free random walks. Close to the critical point we expect the
scaling form
\begin{equation}
N(s,p) = \xi(p)^{d_{\rm f}-3} h_{\pm}(s/\xi(p)^{d_{\rm f}})
\label{h}
\end{equation}
to hold, with scaling functions $h_{+}(x)$ for $p>p_{\rm c}$ and
$h_{-}(x)$ for $p<p_{\rm c}$ that have distinct behaviour at large $x$
but share the same power law form, $h(x)\sim x^{2-\tau}$, 
at small $x$. Since $N(s,p)$ should be independent of $\xi(p)$
at small $s$, we have the exponent relation
\begin{equation}
d_{\rm f} = \frac{3}{\tau - 1}\,.
\label{tau-df}
\end{equation}  
Data at $p=p_{\rm c}$ are shown in the inset to Fig.~\ref{fig4}: they
confirm power law behaviour over six decades in $s$, yielding the
exponent value
\begin{equation}
\tau = 2.184 \pm 0.003
\end{equation}
and hence $d_{\rm f} = 2.534 \pm 0.009$. Using our values for $d_{\rm f}$
and $\nu$, we demonstrate  in the main panel of Fig.~\ref{fig4} scaling collapse
following Eq.~(\ref{h}) for seven values of $p$ in the insulator (0.250, 0.253, 
0.255, 0.256, 0.257, 0.258, and 0.259)
and five values in the metal (0.261, 0.262, 0.263, 0.265, and 0.270).
We note the excellent overlap of data in each phase. Both scaling
functions $h_{\pm}(x)$ follow the power-law behaviour of the critical point at small
$x$. At large $x$, $h_{-}(x)$ falls off exponentially, while 
$h_{+}(x)$ tends to a constant: the escape probability.

These results can be combined with those of
Ref.~\cite{beamond} to conclude for the quantum density of states
that: 
(i) $\rho(\varepsilon,p) \propto \varepsilon^2$ in the insulator;
(ii) $\rho(\varepsilon,p_{\rm c}) \propto |\varepsilon|^{\tau-2}$ at the
critical point; while in the metal  
(iii) $\rho(0,p) \propto (p-p_{\rm c})^{(3-d_{f})\nu}$ and
(iv) $[\rho(\varepsilon,p) - \rho(0,p)] \propto |\varepsilon|^{1/2}$.
For comparison, the $\epsilon$ expansion gives \cite{senthil} at
leading order:
$\rho(\varepsilon,p_{\rm c}) \propto |\varepsilon|^{\epsilon/2}$ and
$\rho(0,p) \propto (p-p_{\rm c})$. 

\begin{figure}
\includegraphics[width=.48\textwidth]{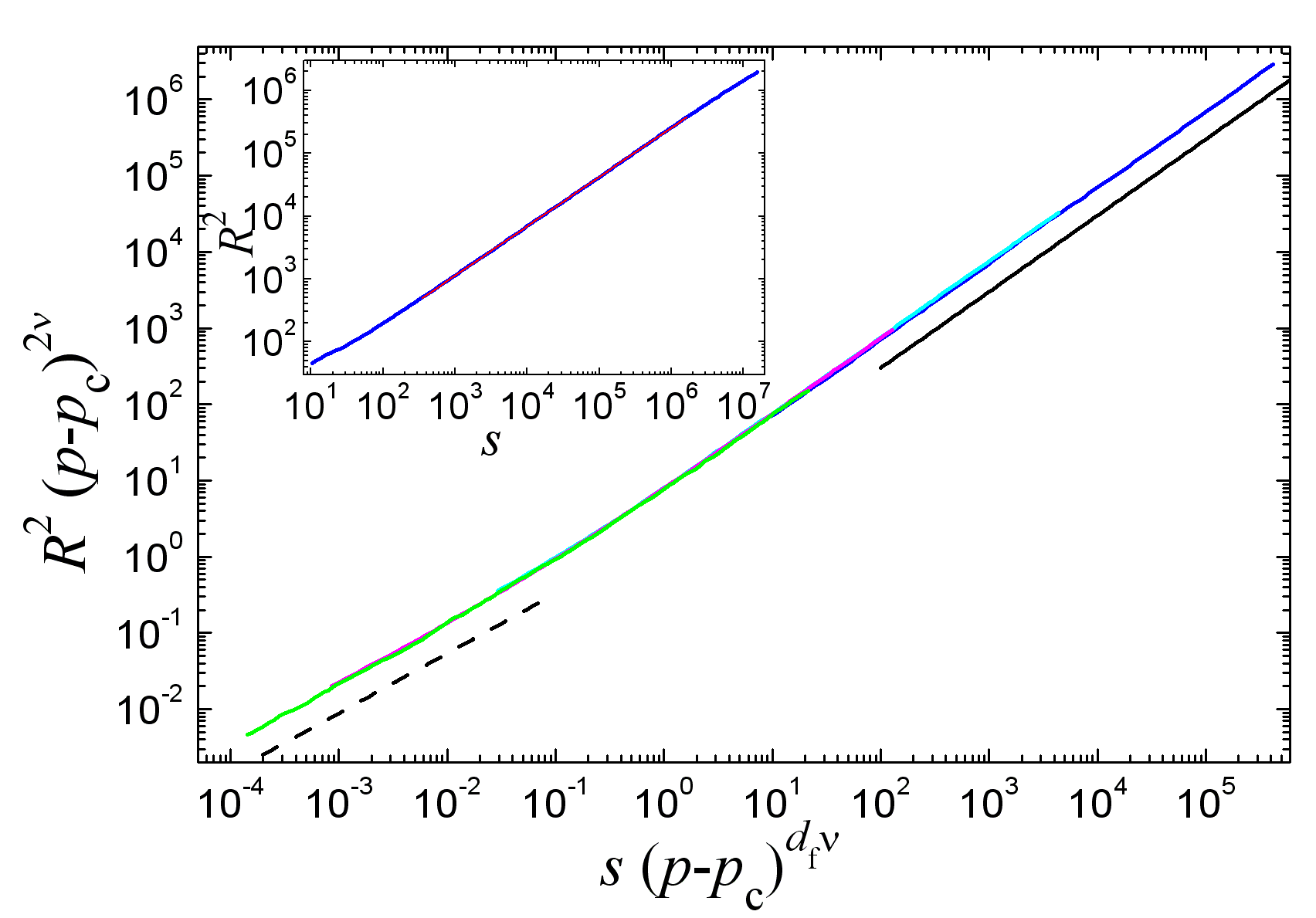}
\caption{(Color online) $\langle R^2(s)\rangle/\xi(p)^2$ as a function of
$s/\xi(p)^{d_{\rm f}}$ in the metal ($p= 0.265$,
0.27, 0.3 and 0.5), on double logarithmic
scales. The continuous and dashed lines have gradients of $1$ and 
$2/d_{\rm f}$, respectively.
Inset: $\langle R(s)^2\rangle$ vs $s$ at the critical point.}
\label{fig5}
\end{figure}

To examine further the properties of classical walks at the critical
point and in the metal we have calculated  $\langle R(s)^2\rangle$ for
very long trajectories, which can be generated individually in
samples of unbounded size. We class a trajectory as long if it has not
closed on itself after $5 \times 10^7$ steps, and we analyse behaviour
for the first $10^7$ steps. At $p=p_{\rm c}$ we expect to see with
this method the fractal behaviour of critical walks, and 
in the inset to Fig.~\ref{fig5} we show that $\langle R(s)^2\rangle$ 
varies as a power of $s$. From this we extract $2/d_{\rm f}=0.784\pm
0.004$, which is consistent with the value determined from $\tau$
using Eq.~(\ref{tau-df}). In the extended phase we expect scaling collapse
of $\langle R(s)^2\rangle/\xi(p)^2$  plotted as a function of
$s/\xi(p)^{d_{\rm f}}$ for different $p$. This is illustrated in the
main panel of Fig.~\ref{fig5} using four values of $p$.
Two power law
regimes are apparent: $\langle R(s)^2\rangle \sim s^{2/d_{\rm f}}$ for
small $s/\xi(p)^{d_{\rm f}}$, as at the critical point; and  $\langle
R(s)^2\rangle \sim s$ for large $s/\xi(p)^{d_{\rm f}}$, as for free
random walks.

\begin{table}[t]
\label{model}  
\begin{tabular}{|c|c|c||c|c|c|}\hline
 Hexagon       & \multicolumn{2}{|c||}{Initial Sites} & Hexagon &
 \multicolumn{2}{|c|}{Initial Sites}\\     \hline\hline
$\ag,-\dg,\bg$ & $(2,0,2)$&$(2,-2,4)$ & $\ag,-\cg,\dg$ &
$(0,0,0)$&$(0,-2,2)$ \\ \hline
$\dg,-\cg,\bg$ & $(4,2,2)$&$(4,4,4)$ & $\ag,-\bg,\cg$ & $(4,2,2)$&$(4,4,4)$\\ \hline
\end{tabular}
\caption{hexagons and their initial sites in the unit cell.}
\end{table}
We close with a full description of our model, deferred
above. Consider a diamond lattice. We take as Cartesian coordinates
for the two sites in the primitive unit cell: ${\bf r} = (0,0,0)$ on
sublattice $A$, and ${\bf r} = (1,1,1)$ on sublattice $B$. At an
$A$-site the position 
vectors to neighbouring $B$-sites are: $\ag = (1,1,1)$, $\bg =
(-1,-1,1)$, $\cg = (-1,1,-1)$ and $\dg = (1,-1,-1)$. For $p=0$ all
classical paths in the model are closed hexagons. They can be specified by giving
an initial site and the first three steps. The unit cell of the model
consists of such eight hexagons, as listed in Table 1. The model does
not retain the full point symmetry of the diamond lattice, being
instead tetragonal. It is, however, not very anisotropic:
the ratio of conductivities in the two distinct directions is about
$1.1$ and depends very little on $p$. 

In conclusion, we have used the mapping to classical walks as a means to investigate
the three-dimensional class C localisation transition with high
precision. Possible future extensions include the study of two and
three particle Green functions \cite{mirlin} and of lattices with
higher coordination number \cite{cardy}. Most importantly, an
analytical understanding of the classical problem, perhaps based on
ideas from polymer physics \cite{polymers}, would be very desirable.

This work was supported by EPSRC Grant No. EP/D050952/1, by DGI
Grant No. FIS2006-11126, and by Fundacion
Seneca, Grant No. 03105/PI/05.


\begin{thebibliography}{99}


\bibitem{review} 
F. Evers and A. D. Mirlin, Rev. Mod. Phys. {\bf 80}, 1355 (2008).

\bibitem{gruzberg}
I. A. Gruzberg, A. W. W. Ludwig, and N. Read, Phys. Rev. Lett. {\bf
  82}, 4524 (1999). 

\bibitem{beamond}
E. J. Beamond, J.  Cardy, and J. T. Chalker, Phys. Rev. B {\bf 65},
214301 (2002).

\bibitem{altland-zirnbauer}
A. Altland and M. R. Zirnbauer, Phys. Rev. B {\bf 55}, 1142 (1997);
M. R. Zirnbauer, J. Math. Phys. {\bf 37}, 4986 (1996). 

\bibitem{senthil}
T. Senthil {\it et al.},
Phys. Rev. Lett. {\bf 81}, 4704 (1998); 
T. Senthil and M. P. A. Fisher, Phys. Rev. B {\bf 60},
6893 (1999). 

\bibitem{zirnbauer}
R. Bundschuh {\it et al.},
Phys. Rev. B {\bf 59}, 4382 (1999);
A. Altland, B. D. Simons, and M. R. Zirnbauer, Phys. Rep. {\bf 359},
283 (2002). 

\bibitem{one-d}
P. W. Brouwer {\it et al.},
Phys. Rev. Lett. {\bf 85}, 1064 (2000); M. Titov {\it et al.},
Phys. Rev. B {\bf 63}, 235318 (2001).

\bibitem{network}
J. T. Chalker and P. D. Coddington, J. Phys. C {\bf 21}, 2665 (1988).

\bibitem{kagalovsky} 
V. Kagalovsky, B. Horovitz, and Y. Avishai, Phys. Rev. B {\bf 55},
7761 (1997); V. Kagalovsky, B. Horovitz, Y. Avishai, and
J. T. Chalker, Phys. Rev. Lett. {\bf 82}, 3516 (1999).

\bibitem{evers}
F. Evers, A. Mildenberger, and A. D. Mirlin, Phys. Rev. B {\bf 64},
241303(R) (2001).

\bibitem{romer}
O.Schenk, M. Bollh\"ofer, and R. A. R\"omer, SIAM J. Sci. Comp. {\bf
  28}, 963 (2006).

\bibitem{markos}
J. Brndiar and P. Markos, Phys. Rev. B {\bf 74}, 153103 (2006).

\bibitem{slevin}
K. Slevin and T. Ohtsuki, Phys. Rev. Lett. {\bf 82}, 382 (1999);
Phys. Rev. B {\bf 63}, 045108 (2001). 

\bibitem{transfer}
J.-L. Pichard and G. Sarma, J. Phys. C {\bf 14}, L127 (1981);
A. MacKinnon and B. Kramer, Z. Phys. {\bf 53}, 1 (1983). 

\bibitem{numerical-recipes}
Ch. 15 {it Numerical Recipes in Fortran}, W. Press, B. Flannery, and 
S. Teukolsky (Cambridge University Press, 1992).

\bibitem{previous}
V. Kagalovsky, B. Horovitz, and Y. Avishai, Phys. Rev. Lett. {\bf 93},
246802 (2004).

\bibitem{andreas} A. W. W. Ludwig, private communication.

\bibitem{mirlin}
A.D. Mirlin, F. Evers, and A. Mildenberger,
J. Phys. A {\bf 36}, 3255 (2003).

\bibitem{cardy}
J. L. Cardy, Comm. Math. Phys. {\bf 258}, 87 (2005).



\bibitem{polymers}
P. G. de Gennes, Phys. Lett. {\bf 38A}, 339 (1972).


\end{thebibliography}
\end{document}